\documentclass[fleqn,10pt]{wlscirep}
\usepackage{amsmath}
\usepackage{float}
\usepackage[utf8]{inputenc}
\usepackage[T1]{fontenc}
\title{The Role of Topology in the Synchronization of Neuronal Networks Based on the Hodgkin-Huxley Model}

\author[1]{Arefeh Mazarei}
\author[2]{Mohammad Amirian Matlob}
\author[3]{Gholamhossein Riazi}
\author[4,*]{Yousef Jamali}

\affil[1,2,4]{Biomathematics Laboratory, Department of Applied Mathematics, Tarbiat Modares University, Tehran, Iran (\url{y.jamali@modares.ac.ir}).}
\affil[3]{Institute of Biochemistry and Biophysics, University of Tehran, Tehran, Iran.}

\begin{abstract}
Complex systems in the real world can be modeled as a network of connected components. The human brain, as a network of neurons among which the interactions cause perception, is a complex network. Synchronization is a dynamical phenomenon that can be seen in the brain. The network topology has a remarkable impact on both the function and the dynamics of neural networks. In this research, synchronization of neural networks is scrutinized through creating various topologies. These networks include both excitatory and inhibitory neurons. We investigate the dynamics of different networks by random rewiring of the synaptic connections. In this manner, a regular network transforms into a small-world network and then becomes a random network. Coherence level which is measured and utilized as the criteria to analyze synchronicity, experiencing a sharp increase as the network changes into the small-world network and growing steadily by the end. On the other hand, a decreasing trend of coherence level is revealed starting from a complete excitatory network and gradually increasing of inhibitory neurons. Thus, the coherence level reaches approximately zero in a complete inhibitory network. By increasing the number of neurons in the network, the degree of synchronization follows a power-law distribution; however, the number of synaptic connections of each neuron and their conductance have a positive impact on synchronization. By applying the model to a C-elegance neural network, not only the mentioned parameters but also the role of degree distribution are highlighted.
\\ \\\textbf{KEYWORDS:} Graph theory, Brain dynamics, Complex networks, Synchronization

\end{abstract}

\begin{document}

\flushbottom
\maketitle
\thispagestyle{empty}

\section{Introduction}

The new observation on the brain’s function not only is known as a means of self-knowledge, but also expresses solidarity between what psychologists, anthropologists, linguists, and philosophers believe.

The importance of networks has been realized in the social sciences, turning networks into a heated issue in the natural sciences, particularly in the study of complex biological systems, including the brain. The strong association between network science and neuroscience helps us to get a better understanding of the structure and function of the brain \cite{sporns2010networks}. A human brain consists of around 100 billion neurons which each neuron forms approximately quadrillion electrical and chemical synaptic connections with other neurons in the brain network \cite{das2014highlighting}. Interactions between complex components of the brain continually create complex patterns. The analysis of the network patterns and connectivity illuminate a number of problems concerning the integration of brain function \cite{sporns2010networks}.

Initial approaches based on the graph theory have attracted researchers to be involved in the understanding the neuronal structure of the brain. Graph theory indeed provides scholars with a simplified and more generalized approach to study the complex neuronal structures such as brain networks. The power of graph-based approaches stems from the point of view that almost all the complex systems can be meaningfully explained as networks \cite{sporns2010networks}. The researches on graph theoretical approaches provide a better insight into information flow and the integration properties of the network\cite{das2014highlighting}.

Graph theory is dealing with the system topology, not with its anatomy \cite{rubinov2010complex}; which is used to abstractly define a nervous system as a set of nodes denoting anatomical regions and interconnecting edges with the structural or functional or effective connections \cite{bullmore2011brain, papo2014complex}. The mean of abstraction of graphs from the details of the underlying data is that the same mathematical language can be employed to determine topological properties compared to random graphs, or graphs obtained from other neuroscience data or non-neural complex systems \cite{bullmore2011brain, papo2014complex}. Topological properties of complex systems can be scrutinized by network science as many of them have already been investigated in brain networks \cite{bullmore2009complex}.

The dilemma of the interplay between structure and function is tractable through the understanding of the relationship between the systems topology and its dynamical processes. A natural way to overcome this challenge in the brain is the simulation of neurons dynamic connected with different topologies. Synchronization is one of the noticeable and commonplace dynamical phenomena in the brain, which can be defined as a coordination between different individuals behavior through coupling them. 

Synchronization among coupled oscillators and other aspects of synchronization such as synchronization stability with a small disturbance have widely been studied in physics and mathematics \cite{barahona2002synchronization, blekhman1988synchronization}. The first studies were associated with the synchronization of periodic systems such as clocks or flashing fireflies \cite{boccaletti2002synchronization,blekhman1988synchronization, pikovsky2003synchronization}. However, much interest has recently been devoted to the synchronization of chaotic systems \cite{boccaletti2002synchronization}, which is widespread in nature, as diverse as lasers \cite{otsuka2000synchronization}, neural networks \cite{hansel1992synchronization}, and many physiological processes \cite{glass2001synchronization}. Among many important works on the synchronization problems of neural networks, in ref \cite{wang2008adaptive} and references cited therein, synchronization of neural networks with or without time-varying delay or distributed delay are discussed.

Based on studies which have highlighted the association of gamma oscillations with attentiveness \cite{bouyer1981fast}, sensory perception \cite{gray1989oscillatory}, cognitive processing \cite{joliot1994human}, memory working \cite{axmacher2007sustained}, learning process \cite{turner2008olfactory}, and movement \cite{murthy1992coherent}, in ref \cite{bazhenov2008effect} in addition to the effect of synaptic connectivity, the network geometry has been indicated, experimentally and computationally, as a critical factor influencing the long-range  synchronization of gamma activity in real biological systems. Ref \cite{aradi2002modulation} also demonstrates the influence of inter-neuronal population variance, the behavior of the neural network, and the synchrony of inter-neuronal firing as well as altering the neurological diseases.

Epilepsy, as a result of the synchronized bursting of neural populations, has been identified by spontaneous recurrent seizures \cite{bogaard2009interaction}. Studies of epilepsy models provide experimental data on the dynamic of brain areas \cite{derchansky2008transition, jin2006enhanced}. Results of ref \cite{bogaard2009interaction} suggest that intrinsic properties of neurons as well as network structure leading to the generation of seizure-like synchrony. However, functional (rather than anatomical) connectivity determined by the spatial pattern of brain trauma (trauma neurons and dominant synaptic connections) is suggested in order to reveal the ways in which a traumatized brain can become epileptic \cite{volman2011topological}. Beside broad studies on macro-synchronized networks resulting in epileptic activity, ref \cite{muldoon2013spatially} focuses on the relationship between structure and dynamics of epileptic networks through studying functional structure at the level of individual neurons. It also highlights the composition of synchronized cell clusters producing network dynamics.

In this research, we try to scrutinize the synchronization concept under different topological conditions of the brain network. Synchronization of a dynamical network will be investigated by simulation of the neural network. Ultimately, the output of the simulation is compared with a real C-elegans network.

\section{Materials \& Methods}
\subsection{Graph Theory}
A graph is an ordered triple $G=( V(G),E(G),I_G)$ comprising a nonempty set $V(G)$, a distinction set $E(G)$ from $V(G)$, and $I_G$ which maps each element of $E(G)$ to an unordered pair of $V(G)$. The elements of $V(G)$ and $E(G)$ are also known as nodes (vertices) and edges of G respectively. 

Clustering coefficient is a measure relating to the degree of nodes which demonstrates how nodes tend to cluster together. Also, this measure indicates the interconnection probability of the node’s neighbors. The average of clustering coefficients of nodes is the clustering coefficient of a network. High clustering coefficient is associated with resilient and flexibility of network against intentional attacks \cite{sporns2010networks}. Short path length is the least number of existing edges between two nodes. The distance between two nodes is the measure of short path length \cite{sporns2010networks}. Centrality is an indicative measure of node’s importance in the network. Also, hubs can be referred to as nodes with high connections playing a significant role in information transmission.

\subsection{Network topology}
Networks, due to their topologies, can be classified in distinct groups. Random network, regular network, small-world network and scale-free network are four networks which have largely been studied so far. Pair nodes of a random network are connected with identical probability, and they have a normal degree distribution with low levels of clustering coefficient. Regular networks have an ordered pattern of connections between nodes, and unlike random networks, they have a much higher clustering coefficient and longer path length. Since the random and regular networks are the kind of idealized models, they are incapable of describing most of the real-world networks. The small-world network (e.g. Watts and Strogatz (WS) model) has a large amount of clustering, and unlike random networks, they have a shorter path length. In fact, there is a balance between segregation and integration. In the scale-free networks, there is a few highly connected nodes and plenty of nodes with low degree. That is, the main features of scale-free networks are extremely extensive and nonhomogeneous degree distributions indicating the low efficiency of the network \cite{sporns2010networks, das2014highlighting}.

\subsection{Biological models of neuron}
Neuron is the fundamental unit of the neural system \cite{hall2015guyton}. Neurons play a key role in data processing of the brain, and membrane potential is the most important physical variable in this process. Each neuron responds to the stimulus by disorienting membrane ionic arrangement and leaving the neuron out of the rest state. A biological model of the neuron is a mathematical description of neuronal features which is designed to accurately describe and predict biological processes. Many models have been proposed to investigate the dynamic behavior of neurons, including Artificial neuron model, Integrate \& fire model \cite{abbott1999lapicque}, Leaky integrate \& fire model \cite{koch1998methods}, Exponential integrate \& fire model \cite{badel2008dynamic}, Hodgkin-Huxley model \cite{hodgkin1952quantitative}, Fitzhugh-Nagumo model \cite{izhikevich2006fitzhugh}, Morris-Lecar model \cite{morris1981voltage}, and Hindmarsh-Rose model \cite{hindmarsh1984model}.

\begin{figure}[ht]
\centering
\includegraphics[scale=0.45]{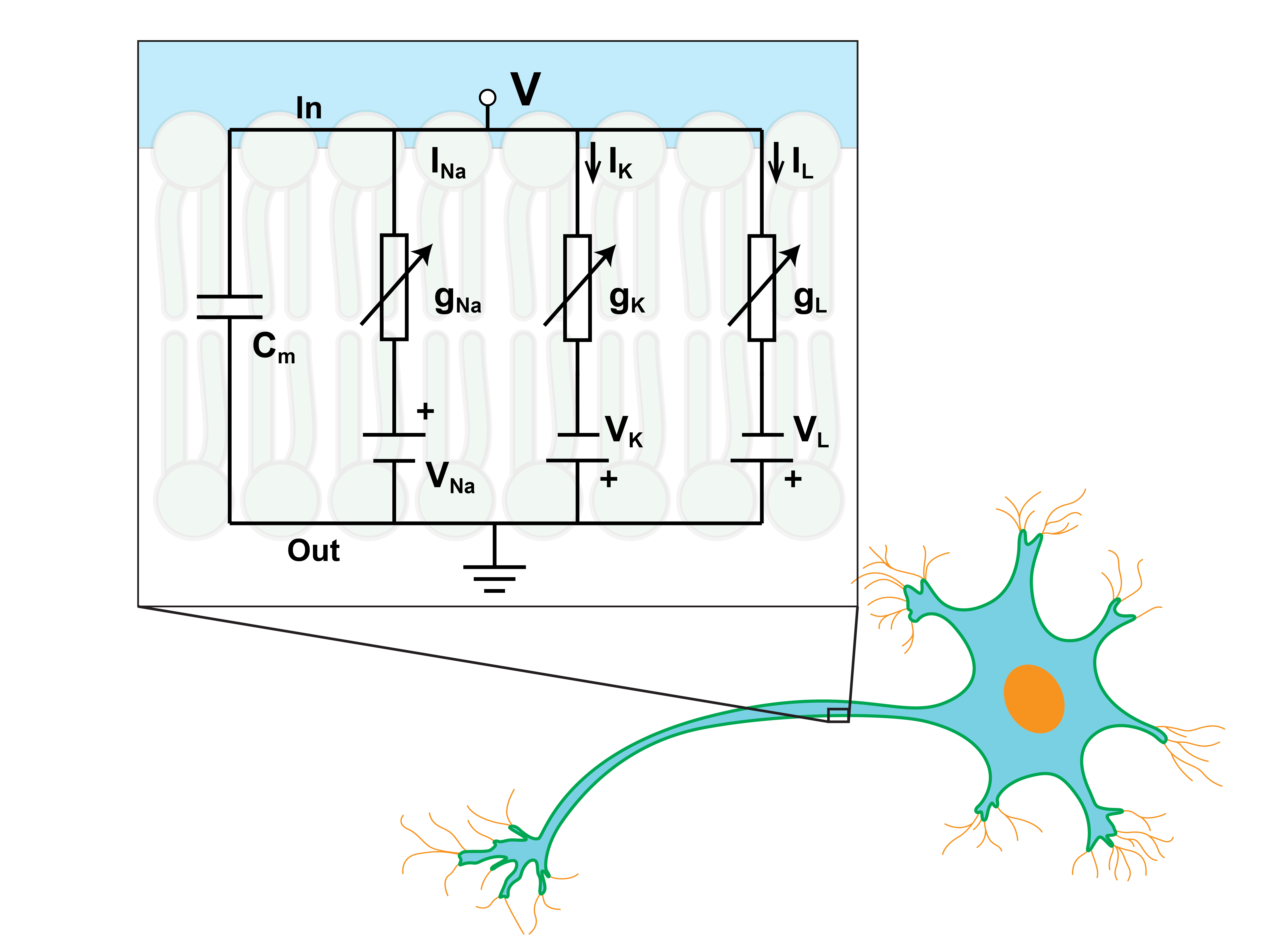}
\caption{Cell and current model in a simple circuit.}
\label{fig:1}
\end{figure}

One of the most successful models of neurons is Hodgkin–Huxley model developed by Hodgkin and Huxley in 1952 to explain the ionic mechanisms in the squid giant axon. This model is based on a simple circuit composed of batteries, resistors, capacitors and multiple currents related to voltage (figure (\ref{fig:1})) \cite{siciliano2012hodgkin}. The current can pass the circuit through ion channels in the membrane or by charging the capacitor of the membrane. As an advantage of this model, we can determine the membrane’s capacity independently without considering the sign and size of the intracellular potential with the least impact of time \cite{hodgkin1952quantitative}. It is assumed that the capacitance of the membrane ($C$) is constant and the passing voltage ($V$) through the membrane changes with total current ($I_{tot}$). $I_{Na}$, $I_K$ and $I_L$  are the currents passing through sodium, potassium and leakage channels respectively and $I_{ext}$ represents incoming current from the external source. These currents can be calculated with Ohm’s law. $g$ is the conductance or the reverse of the resistance.  $m, n$ and $h$ are three variables associating with the probability of activation or inactivation of the gates related to sodium and potassium channels; also $\alpha$ and $\beta$, alternatively, are opening or closing rate of each gate deduced from empirical measurements. In summary, the total passing current of one neuron can be described by the following equations.

\begin{align}\label{eq:1}
C\dfrac{{dV(t)}}{{dt}} = \sum\limits_i {{I_i}(t,V) = } {I_{ext}} + {I_{Na}} + {I_K} + {I_L} = {I_{ext}} - {g_{Na}}{m^3}h(V - {V_{Na}}) - {g_K}{n^4}(V - {V_K}) - g_{L}(V - {V_L}),
\end{align}
\begin{align}\label{eq:2}
  \dfrac{{dm}}{{dt}} &= {\alpha _m}(V)(1 - m) - {\beta _m}(V)m, \\ 
  \dfrac{{dh}}{{dt}} &= {\alpha _h}(V)(1 - h) - {\beta _h}(V)h,  \\\label{eq:3}
  \dfrac{{dn}}{{dt}} &= {\alpha _n}(V)(1 - n) - {\beta _n}(V)n, \\\label{eq:4}
  \nonumber
\end{align}
\begin{align}
{\alpha _m}(V) &= 0.1(V + 40)/(1 - \exp ( - (V + 40)/10)), \\ \label{eq:5}
  {\beta _m}(V) &= 4\exp ( - (V + 65)/18),  \\ \label{eq:6}
  {\alpha _h}(V) &= 0.07\exp ( - (V + 65)/20), \\ \label{eq:7}
  {\beta _h}(V) &= 1/(1 + \exp ( - (V + 35)/10)), \\ \label{eq:8}
  {\alpha _n}(V) &= 0.01(V + 55)/(1 - \exp ( - (V + 55)/10)), \\ \label{eq:9}
  {\beta _n}(V) &= 0.125\exp ( - (V + 65)/80). \\ \label{eq:10}
  \nonumber
\end{align}

\subsection{Biological models of synapse}
In the nervous system, neuronal functions cannot be sustained by individual neurons, and it is essential to be connected to each other. This connection at the neuronal conjunction is called synapse. Synaptic transmissions, due to its complexity, can be investigated by various approaches.

The response of a neuron to a neurotransmitter can be modeled based on one form of Hodgkin-Huxley model. As a modified model, Synaptic current ( $I_{syn}$) is
\begin{equation} \label{eq:11}
I_{syn}= g_{syn} [O](V-E_{syn}).    
\end{equation}
Here $g_{syn}$ is synaptic conductance, $E_{syn}$ is reversal potential for one receptor, $V$ is membrane potential, $[O]$ is synaptic gating variable; on the other words,  $[O]$ is the percentage of open channels, and $[T]$ is the concentration of neurotransmitters. The changes of the percentage of open channels can be represented by the following linear differential equation.
\begin{equation} \label{eq:12}
    \frac{d[O]}{dt}= \alpha[T](1-[O])- \beta[O].
\end{equation}
Due to the type of receptors, synaptic dynamics can be described by following set of equations \cite{destexhe1998kinetic}.
\begin{align}\label{eq:13}
  {I_{AMPA}}(t,V) &= {{\bar g}_{AMPA}}.[O].(V(t) - {E_{AMPA}}), \\ 
  {I_{NMDA}}(t,V) &= {{\bar g}_{NMDA}}.B(V).[O].(V(t) - {E_{NMDA}}), \\ \label{eq:14}
  {I_{GABA_A}}(t,V)&= {{\bar g}_{GABA_A}}.[O].(V(t) - {E_{Cl}}),  \\   \label{eq:15}
  {I_{GABA_B}}(t,V)&= {{\bar g}_{GABA_B}}.\frac{{{{[G]}^n}}}{{{{[G]}^n} - {k_d}}}.(V(t) - {E_K}), \\ \label{eq:16}
  \nonumber
\end{align}
\begin{align} 
  [T]({V_{pre}}) = \dfrac{{{T_{\max }}}}{{1 + \exp ( - ({V_{pre}} - {v_p})/{k_p})}},\\\label{eq:17}
  \nonumber
  \end{align}
\begin{align}
  B(V) &= 1/(1 + \exp ( - 0.062V)[Mg^{2+} ]/3.57), \\ \label{eq:18}
  \dfrac{{d[G]}}{{dt}} &= {k_1}[O] - {k_2}[G]. \\ \label{eq:19}
  \nonumber
\end{align}
Where $V, \overline{g} ,E, [O], [T],B(V),[G]$ are postsynaptic voltage, the maximum conductance,  equilibrium potential or reversal potential, the percentage of open receptors, the concentration of neurotransmitters, Magnesium block and the concentration of G-protein, respectively. $k_d$ is also the segregation constant.

\subsection{Synchronization}
Synchronization, an observable dynamical phenomenon in the brain, can be defined as an adaptation of coupled elements’ activities. The study of synchronization on various complex networks from different frameworks illustrates the importance of topology in determining the dynamic behavior of the system. Synchronization among the coupled oscillators and other aspects of synchronization such as stability with a small disturbance have been widely studied in physics and mathematics \cite{barahona2002synchronization, pikovsky2003synchronization, d2008synchronization, wu2008synchronization}. It is possible that synchronization takes place between two elements or even a complex network of elements such as a neural network. It is indeed possible that one part of the system experiences synchronization and the other part does not.

In order to investigate the effect of network structure on synchronization, we first assume that \(N\) oscillators interact in a network with a particular pattern. A characteristic like \(\phi_i (i=1,2,...,N)\) is attributed to every oscillator interacted together due to their interior dynamic and coupling. Dynamical evolution of the system is defined as follow:
\begin{equation}\label{eq:20}
    \frac{d\phi_i}{dt}= f_i (\lbrace \phi \rbrace).
\end{equation}
Where $\lbrace \phi \rbrace$ includes all states. Every isolated oscillator can experiences a stable fixed point, a limit cycle, or a chaotic attractor over the evolution period. The set of all oscillators with the evolution equation (\ref{eq:20}) can be displayed as a network that each node is a representation of an oscillator with connected edges if there is any dynamically connection between the nodes. When a large number of oscillators couple together, various types of synchronization will be visible. Complete synchronization appears in identical chaotic systems. A system is in complete synchronization mode if it identically evolves with a set of initial conditions over the time \cite{qing2005transition}. 

The degree of synchronization in a network including a large number of neurons can be indicated by a number normalized between zero and one. This number so-called measure of synchronization equals zero when the network is asynchronous and equals one when the network completely synchronizes. That is, the system attracts to one state. The other amounts between zero and one demonstrate the local synchronization. Cross-correlation (c-c) is a measure that is defined by a comparison among the activity of two neurons with various lag. c-c can be formulated as: 
\begin{equation}\label{eq:21}
c_{ij}=  \frac{1}{T_m}  \int_0^{T_{m}} x_i(t) x_j (t+\tau), \end{equation}
here $x_i(t)$ is the autonomous activity rate of $i^{th}$ neuron.

Spikes' synchronization is a measure to describe the dynamical state of a neural network. According to normalized c-c among coupled neurons, there is a measure of synchronization to define the level of firing coherence in the network. To do this, we divide the time period $T$ into $\Delta t= \tau$ intervals; in such a way, there is at most one spike within each interval. For each pairs of neurons, there are two 0-1 sequences $\lbrace x_i \rbrace$ and $\lbrace y_i \rbrace$; $i=1,2,...,N_b$  $({T}/{N_b}=\tau)$, and the coherence between them, according to c-c with zero lag, can be calculated as follows:
\begin{equation}\label{eq:22}
    K_{xy}=  \frac{\Sigma_i^{N_b} x_i y_i }{\sqrt{\Sigma_i^{N_b} x_i  \Sigma_i^{N_b} y_i}}.
\end{equation}
Here, $N_b$ represents the number of intervals in each sequence. By calculating the average on the whole network, the result will be a measure between zero and one which describes the level of synchronization of the oscillators.

\subsection{Neuronal network }
Interaction among neurons through synapses causes emergence phenomena in the brain as a complex system. As a result, we can model the brain as a network in which the neurons are the nodes and the synapses are the edges. By doing so, we have designed a network of $N$ neurons whose interaction pattern is determined by the topology of the network. Our network consists of both excitatory and inhibitory neurons. Due to the success of the \textit{Hodgkin–Huxley} model in modern neural biology, we have applied this model to investigate the dynamic of each neuron. According to our model, the total current inside of each neuron is defined as a summation of passing currents from sodium channel, potassium channel, and current from the external source. All of these currents followed the channel kinetics formulated by \textit{Destex \& Pare}. The dynamic of membrane potential and the total current are described by equations (\ref{eq:1}), (\ref{eq:2}),..., and (\ref{eq:9}). The initial values of using parameters are listed in Table (\ref{tab1}). It is also worth noting that we have applied a small random disturbance of 10$msec$ as an external current to each neuron. The resting potential is also considered -65$mV$. The initial value of the potential for each neuron is set randomly with an average of -64$mV$.

\begin{table}[ht]
\centering
\begin{tabular}{|l|l|l|l|l|l|l|l|}
\hline
Parameters  & $g_{Na}$ & $V_{Na}$ & $g_K$ & $V_K$ & $g_L$ & $V_L$ & $C$\\
\hline
Values  & 120 & 55 & 36 & -72 & 0.3 & -49.4 & 1\\
\hline
Units  & $m$S & $m$V & $m$S & $m$V & $m$S & $m$V & $\mu$F \\
\hline
\end{tabular}
\caption{\label{tab1}The initial values of used parameters in the \textit{Hodgkin–Huxley} model \cite{diehl2004modelling}. Also, the initial values of $m, n$, and $V$ are set randomly between zero and one.}
\end{table}
\subsection{Synaptic currents}
The components of excitatory and inhibitory synaptic currents are considered only AMPA receptors and $GABA_A$ receptors, respectively. Dynamic of synaptic currents are defined according to the equations (\ref{eq:12}), (\ref{eq:13}), and (\ref{eq:16}) for excitatory currents and equations (\ref{eq:12}), (\ref{eq:14}), and (\ref{eq:16}) for inhibitory currents. In these equations, $V$ is postsynaptic voltage without any lag in potential transferring. The initial values of parameters are listed in Tables (\ref{tab2}) and (\ref{tab3}). It should be noted that synaptic conductance is set with a constant amount of 0.5$m$S.

\begin{table}[ht]
\centering
\begin{tabular}{|l|l|l|l|l|l|l|}
\hline
Parameters  & $K_p$ & $V_p$ & $T_{max}$ & $\beta$ & $\alpha$ & $E_{AMPA}$\\
\hline
Values  & 5 & 2 & 0.001 & 0.19 & $1.1\times10^3$ & 0\\
\hline
Units  & $m$V & $m$V & $M$ & $msec^{-1}$ & $M^{-1}msec^{-1}$ & $m$V \\
\hline
\end{tabular}
\caption{\label{tab2} The initial values of used parameters in excitatory synaptic currents with AMPA receptors \cite{destexhe1998kinetic}.}
\end{table}

\begin{table}[ht]
\centering
\begin{tabular}{|l|l|l|l|l|l|l|}
\hline
Parameters  & $K_p$ & $V_p$ & $T_{max}$ & $\beta$ & $\alpha$ & $E_{GABA_A}$\\
\hline
Values  & 5 & 2 & 0.001 & 0.19 & $5\times10^3$ & -80\\
\hline
Units  & $m$V & $m$V & $M$ & $msec^{-1}$ & $M^{-1}msec^{-1}$ & $m$V\\
\hline
\end{tabular}
\caption{\label{tab3}The  initial values of using parameters in inhibitory synaptic currents with $GABA_A$ receptors \cite{destexhe1998kinetic}.   }
\end{table}

\subsection{Network structure}
In this work, the networks are constructed based on WS model: A network of $N$ vertices and the average degree of k are defined as a k-regular graph in which for each vertex, $k/2$ of its edges are connected to right vertices and the rest of them  are connected to the left vertices. We have randomly rewired the synaptic connections. In effect, each synapse is chosen and rewired with a probability of p to another randomly chosen neuron. The probability of p has defined the topology of the network. For $p=0$, the network remains regular and for $p=1$, all synapses are rewired and also, the resultant network is a psudo-random network. In the middle, however, a small amount of p leads to a small numbers of long-range synapses while the rest of them are short-range showing the feature of small-world networks. As a result, by changing the amount of p, the topology of the network will be changed from the regular to the small-world network and eventually to the random network. It is reported that the brain network is a small-world, so we have focused on this architecture of networks and have compared the results with regular and random models.

\subsection{Synchronization Analyses Based on Spike Data}
The analysis of the neural activity of neurons will be possible by calculating the spike sequence of neuronal firing. A spike is defined as any high-threshold voltage. Collected data within a span of 1000$msec$ are saved, and according to the synchronization definition, i.e. equation (\ref{eq:22}), the dynamical state of the network is defined. In this manner, we divide the spike sequence into 2$msec$ subintervals containing one spike at most in order to gain a 0-1 sequence. Then, the synchronization criterion of the network is defined as the average coherence measure of pair neurons. We also calculate the standard deviation by repeating the process five times.

\section{Results}
\subsection{Synchronization in neural activity of networks with various topology}
\begin{figure}[ht]
\centering
\includegraphics[scale=0.8]{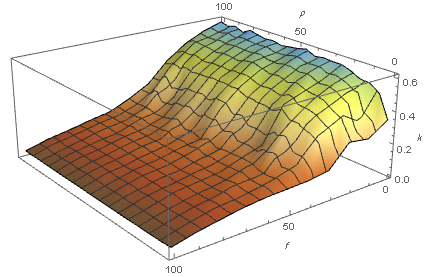}
\caption{Investigation of the spike coherence by changing the network topology from a regular network to a random network, and from a completely excitatory to a completely inhibitory network. $f$ is the percentage of inhibitory neurons in the network. For $f=0\%$ and $f=100\%$, we have a completely excitatory and  a completely inhibitory network respectively. $p$ indicates the percentage of rewiring connections. For p=0\%, there is a regular network, and for p=100\%, we have also a random network. As it is clarified, for about p=5\%, we have a small-world network. The third axis associated with coherence level shows the synchronization index of the network, a number between 0 and 1. The least amount of coherence is related to a regular network, and then it has sharply increased up to the small-world topology. The coherence level fluctuates slightly over the rest of p changes until it represents a random network. On the other hand, increasing the percentage of inhibitory neurons causes a descending tendency of coherence. And with regards to a completely inhibitory network, this index has an almost zero amount regardless of the topology of the network.}
\label{fig:2}
\end{figure}

We have conducted a numerical simulation by designing a network of 100 neurons with the mean degree of 16 according to the dynamics of neuron and synapse defined in the previous section. To investigate the effect of topology on the synchronization dynamics, a regular network of excitatory neurons is defined. By changing p in the range of 0 to 1, the topology of the network has transformed from a regular network to a small-world network and finally led to a sudo-random network. Generally, the results illustrate an ascending trend in synchronization measure as topology is changing. Alteration of the topology has heavily influenced the coherence level. According to the figure (\ref{fig:2}), the minimum amount of coherence is related to a regular network. In the case of the small-world network, the coherence level experiences a sharp increase. Due to the small-world network’s topology, the conclusion seems logical since neural activity spread over an expansive region through the connections of short path length. This upward trend slightly continues while p increases and reaches its peak at $p=1$, which is a random network. We notice that the maximum amount of coherence level is reported for a small-world network in the special conditions with a particular range of synaptic strength \cite{kitano2007variability}.

In addition to the network topology, we have mixed excitatory AMPA neurons with inhibitory $GABA_A$ neurons. We have a complete excitatory network at first and have gradually exchanged a percentage of excitatory neurons to inhibitory ones, thereby having a complete inhibitory network at the end. As the number of inhibitory neurons increases, synchronization index experiences a downward trend falling sharply at first and then continuing with a slight decrease to reach a thoroughly inhibitory network. As illustrated in the figure (\ref{fig:2}), synchronization is almost zero in an inhibitory network, regardless of the topology of the network. As a result, it can quickly be deduced that in an inhibitory network, topology does not have a conspicuous role in the synchronization level. In other words, the excitatory connections characterize the network behavior.

\begin{figure}[H]
\centering
\includegraphics[width=\linewidth]{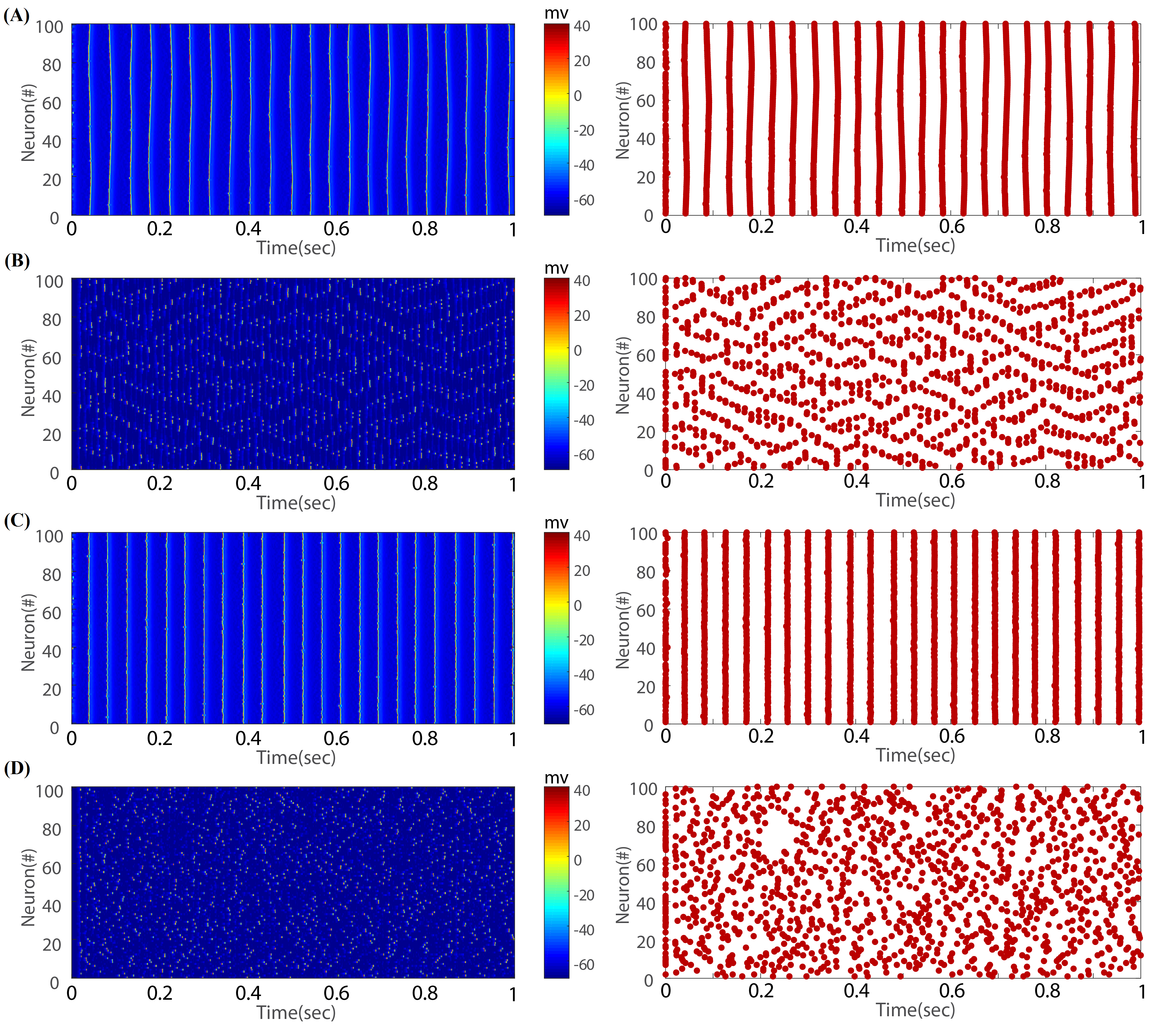}
\caption{Patterns of neuronal firing and potential changes in (A) a regular and completely excitatory network ($p=0,f=0$); (B) a regular and completely inhibitory network ($p=0,f=100$); (C) a random and completely excitatory network ($p=100,f=0$); (D) a random and completely inhibitory network ($p=100,f=100$). For each part, the left figure illustrates the potential changes of each neuron by color spectrum, and the right figure indicates the firing pattern of each neuron during the time. According to patterns, there are synchronous clusters in the excitatory regular network. In the inhibitory regular network, although the potential is transferring from one cluster to another, there is noticeable changes in firing patterns compared to the excitatory regular network. In the excitatory random network, synchronization is observed among a wide range of neurons fully conforming to the high amount of coherence index in figure (\ref{fig:2}). No synchronous pattern in the neuronal firing of the inhibitory random network corresponds with the low amount of coherence level in this type of network. }
\label{fig:3}
\end{figure}

In order to validate our results, we have investigated the dynamical behavior of the system over time. By doing this, the potential changes of neurons have been recorded in four extreme states of  $p$ and $f$ (a regular and completely excitatory network, a regular and completely inhibitory network, a random and completely excitatory network, and a random and completely inhibitory network). According to the figure (\ref{fig:3}-A), synchronization in the excitatory regular network is indicated as clusters of spikes. The rationale behind this fact can be associated with the highly clustered structure of the topology causing neural activities to release slightly through the local connections in neighboring regions and from one cluster to another one. In the inhibitory regular network, figure (\ref{fig:3}-B), synchronization pattern changes extensively and the coherence of clustered spikes is not conspicuous. According to figure (\ref{fig:3}-C), in the excitatory random network, synchronization is observed among a wide range of neurons fully conforming to the high amount of coherence index in the figure (\ref{fig:2}). Obviously, there is not any synchronization pattern among neuronal activities in the random inhibitory network, figure (\ref{fig:3}-D). This reduction of synchronization in inhibitory networks would be associated with the dynamic of inhibitory synapses in the absence of time lag.

\subsection{The influence of significant parameters on synchronization}
To investigate the sensitivity of synchronization dynamic to some of the important parameters defined in the algorithm, including the number of neurons, the number of synaptic connections, and the amount of synaptic conductance, we have examined them on a regular and completely excitatory network.
\subsubsection{Number of neurons in a network}
\begin{figure}
\centering
\includegraphics[scale=0.4]{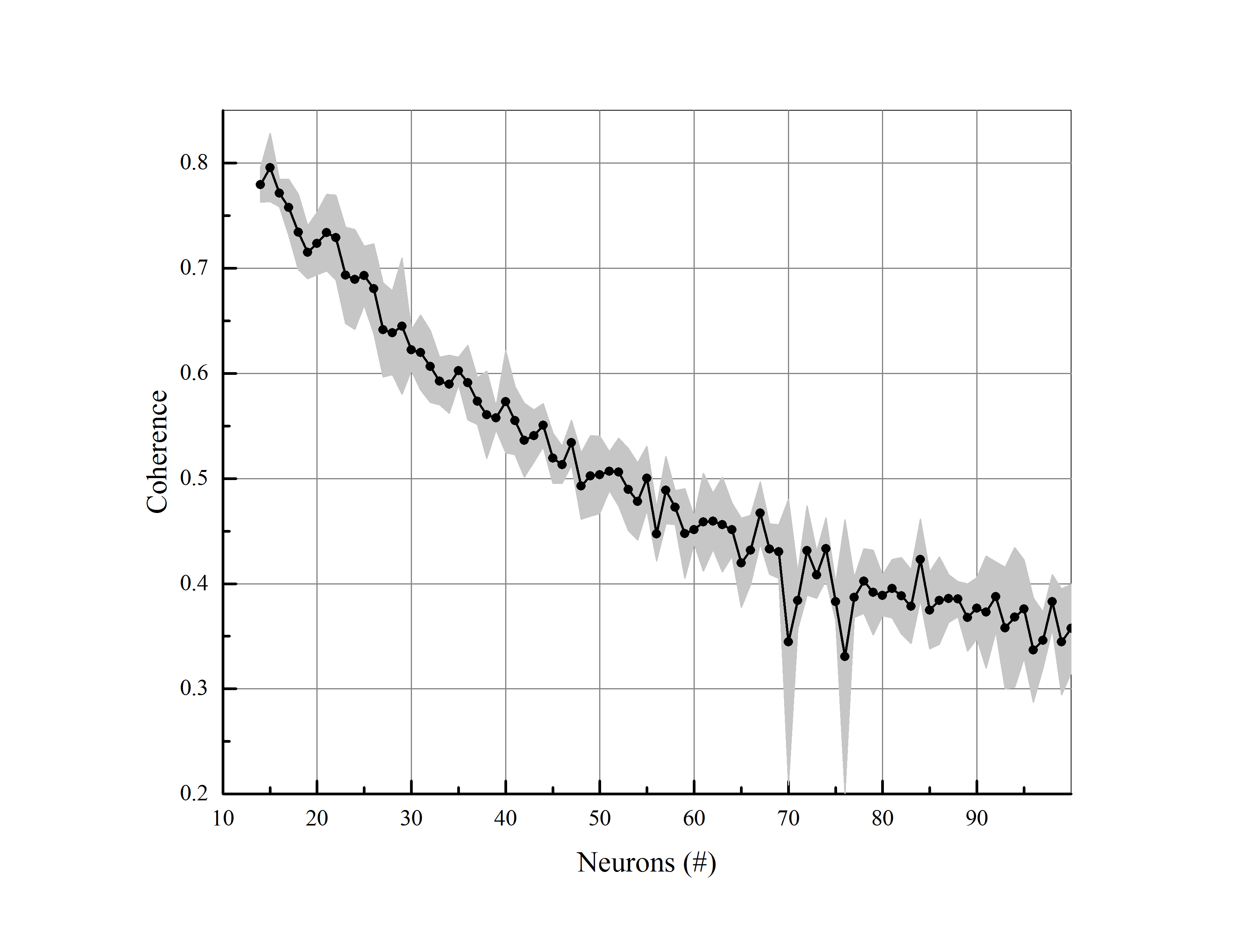}
\caption{The sensitivity of synchronization to the number of neurons. The horizontal axis shows the number of neurons and the vertical axis shows the synchronization index in an excitatory regular network ($p=0,f=0$). The other parameters, especially the number of synaptic connections and synaptic conductance, are fixed. The more neurons, the less synchronization index and consequently the less coherent network. The gray margins show the standard error of the experiment for five-time repetitions.}
\label{fig:4}
\end{figure}

One of the investigated parameters is the number of neurons in the network. As we have increased the number of neurons in an excitatory regular network, while the other parameters especially the number of synaptic connections and synaptic conductance are fixed, the coherence level decreases following a power-law distribution. According to figure (\ref{fig:4}), the coherence level approximates to zero for the enormous number of neurons. In other words, the output of increasing the number of neurons results in the reduction of synchronization. The derived result makes sense because in comparison to a friendship network, the more members the less coherent group; consider the number of connections and the amount of intimacy fixed among the group. On the other hand, increasing the number of neurons results in the decline of the small-world features of the network consisting of the clustering coefficient and short path length. 
\begin{figure}
\centering
\includegraphics[scale=0.4]{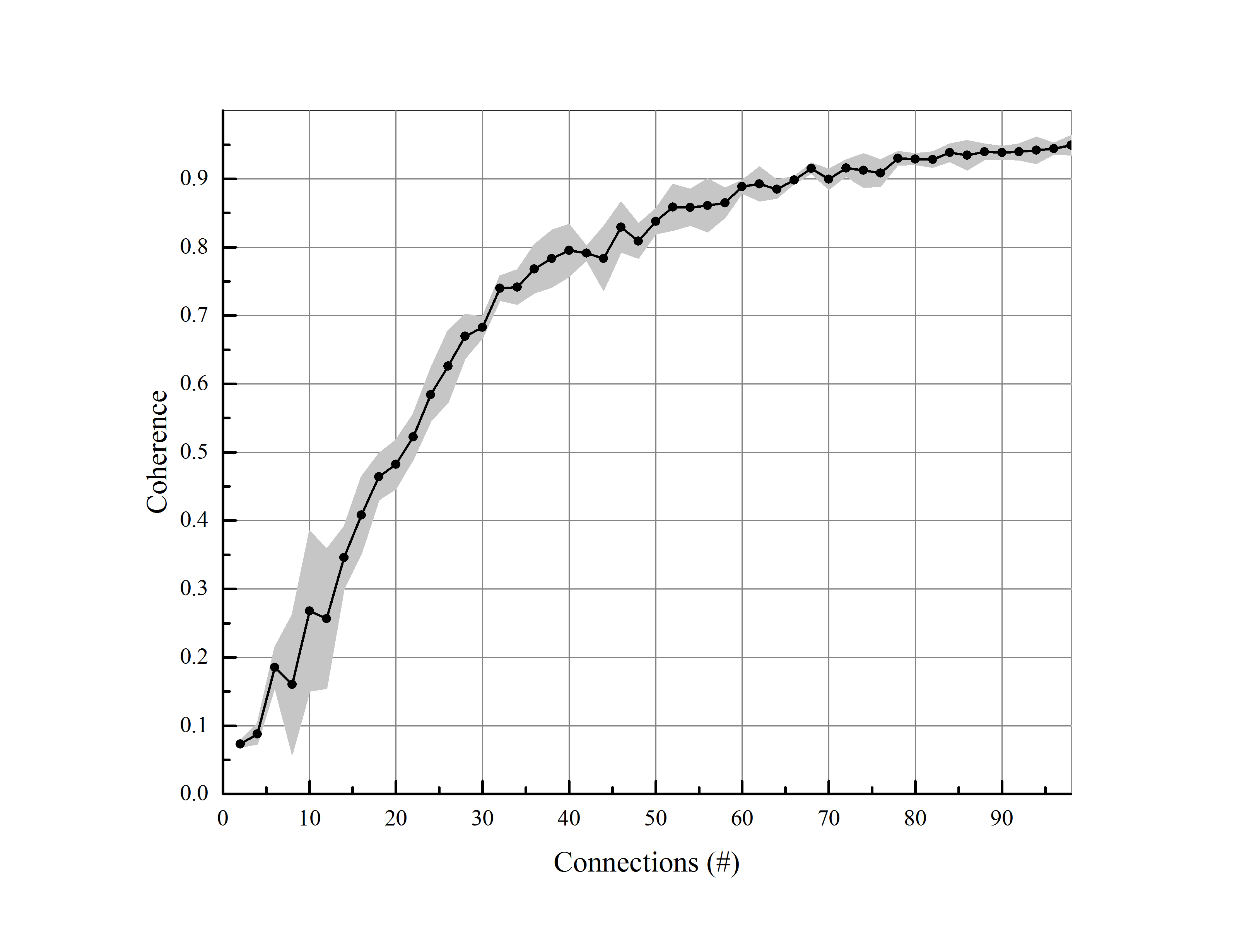}
\caption{The sensitivity of synchronization to the number of synaptic connections. The horizontal axis shows the average connections of each neuron and the vertical axis shows the synchronization index in an excitatory regular network ($p=0,f=0$). It is assumed that the number of neurons and the synaptic conductance are fixed. The more synapses, the more synchronization index and consequently, the higher coherent network. The gray margin concerns with the standard error of the experiment for five-time repetitions.}
\label{fig:5}
\end{figure}
\subsubsection{The number of synaptic connections}
In this part, the number of neurons and the amount of synaptic conductance are remained fixed and also, the number of synapses has changed to a regular and completely excitatory network. As the number of connections increased, synchronization rises with an ascending trend and when all the possible connections exist, for an all-to-all network, the synchronization index will be nearly one. According to figure (\ref{fig:5}), this upward trend is not indeed linear, but logarithmic. In comparison to a friendship network, the more friendship connection among members, the more coherent group; the members and the intimacy degree among them are assumed fix. According to the logarithmic trend, the coherence within a narrow interval increases dramatically. As a result, the least number of connections cause a high level of coherence which is 30 for this experiment.

\subsubsection{Synaptic conductance}
\begin{figure}[H]
\centering
\includegraphics[scale=0.4]{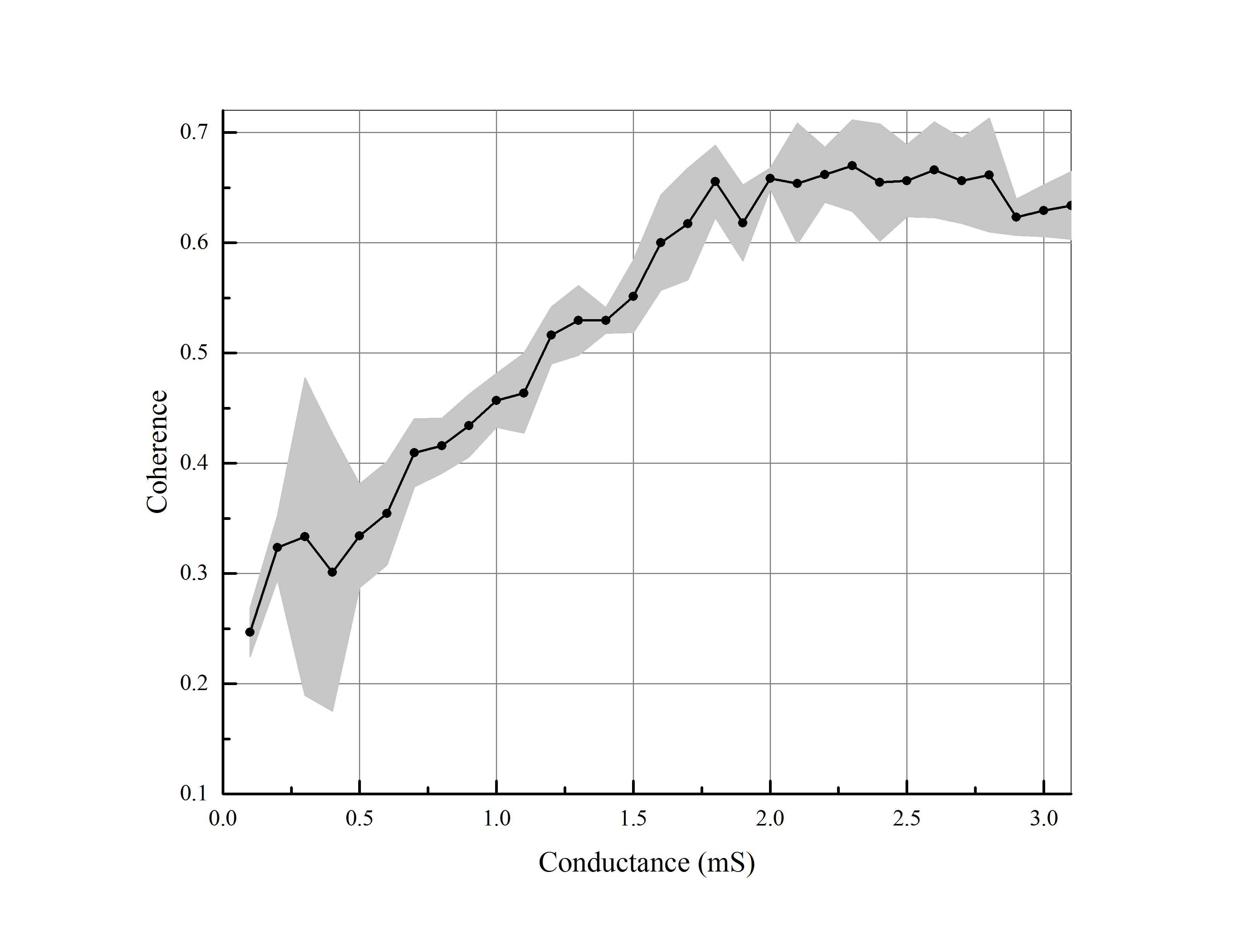}
\caption{The sensitivity of synchronization to synaptic conductance. The horizontal axis shows the amount of conductance for each synapse and the vertical axis shows the synchronization index in an excitatory regular network ($p=0,f=0$). The number of neurons and the number of synaptic connections are fixed. The more synaptic conductance, the more synchronization index and consequently, the higher coherent network. The gray margin shows the standard error of the experiment for five-time repetitions.}
\label{fig6}
\end{figure}

The third investigated parameter is synaptic conductance. In order to have a meaningful cortical potential, we have changed the amount of synaptic conductance to the range of $[10^{-4},2.5]$. Notice that the number of neurons and the number of synaptic connections are fixed in an excitatory regular network. The results illustrate the positive effect of this parameter on synchronization. The conductance can be expressed as synaptic strength or coupling strength. The synaptic strength is the influence of pre-synaptic spike on the post-synaptic potential. Therefore, the more synaptic strength, the more synchronization in the neuronal network. According to figure (\ref{fig6}), the dynamic of synchronization rises almost linearly at first, and then it satiates taking the maximum amount.

\subsection{The largest Eigen-value of coherence matrix}
\begin{figure}[H]
\centering
\includegraphics[width=\linewidth]{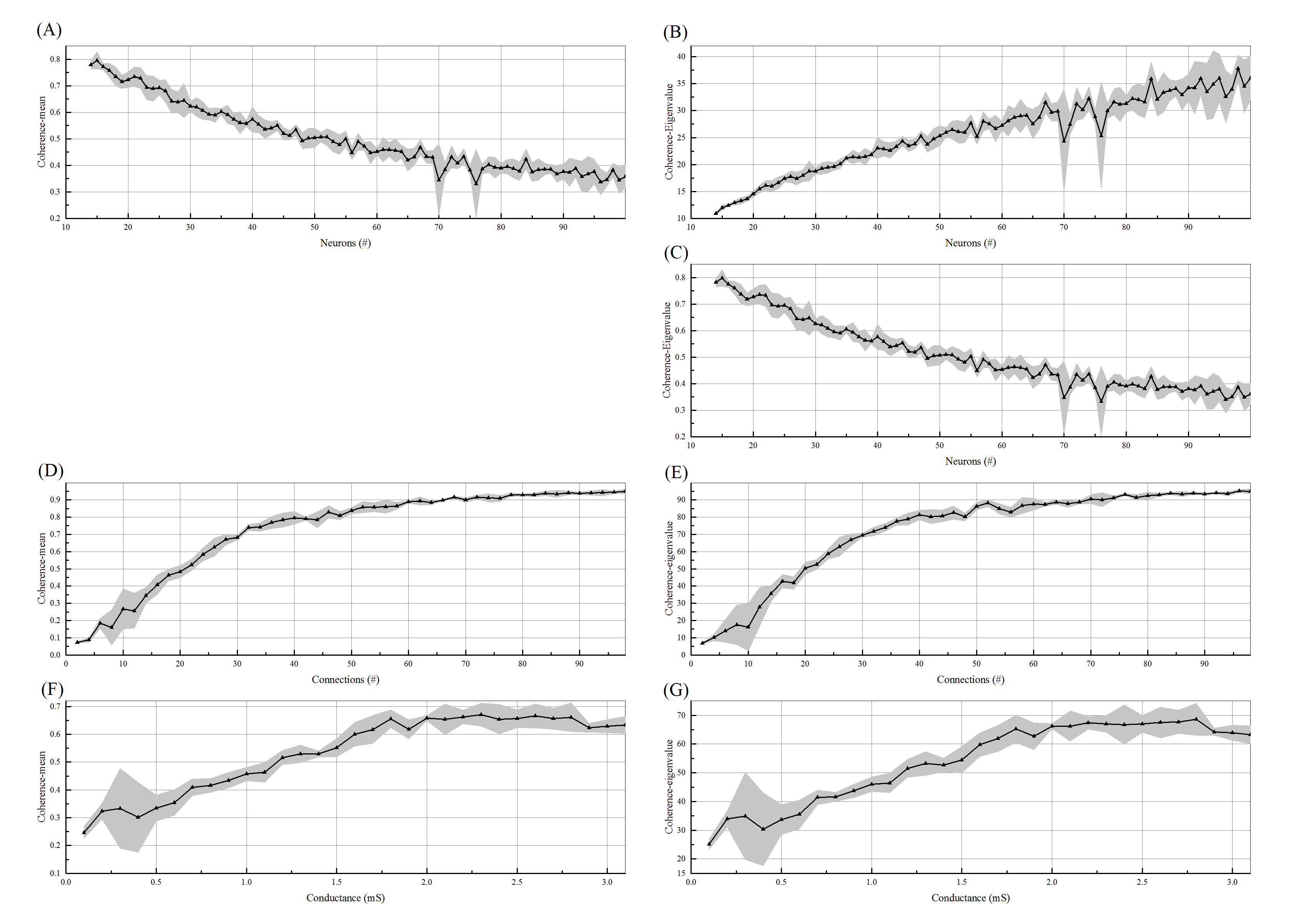}
\caption{Comparison of the changes in synchronization by two definitions, the average and the largest Eigen-value of coherence matrix. (A) shows the effect of the number of neurons on synchronization by defining the average of coherence matrix as synchronization index, (B) expresses the effect of the number of neurons on synchronization by defining the largest Eigen-value of the coherence matrix as synchronization index, (C) concerns with normalization the result of part B, (D) the effect of the number of connections on synchronization by defining the average of the coherence matrix as a synchronization index, (E) the effect of the number of connections on synchronization by defining the largest Eigen-value of the coherence matrix as a synchronization index, (F) the effect of synaptic conductance on synchronization by defining the average of the coherence matrix as a synchronization index, (G) the effect of synaptic conductance on synchronization by defining the largest Eigen-value of coherence matrix as synchronization index. The gray margins show the standard error of experiment for five-time repetitions.}
\label{fig7}
\end{figure}

In order to validate the conducted results in the previous section, we have defined the largest Eigen-value of coherence matrix as another measure for synchronization and examined above parameters. The figure (\ref{fig7}) provides a comparison between the levels of coherence defined by both measures according to the changes of the mentioned parameters. The changes in the coherence level associated with the number of synapses and the amount of synaptic conductance are conspicuously the same. By increasing the number of neurons, the coherence is varying approximately the same but in opposite directions. Therefore, we have normalized the coherence level, achieved from the second measure, by dividing it into the number of neurons. It can clearly be seen that the result is the same as the former measure (the average of coherence matrix).

\subsection{Synchronization in C-elegance network}
\begin{figure}[H]
\centering
\includegraphics[scale=0.45]{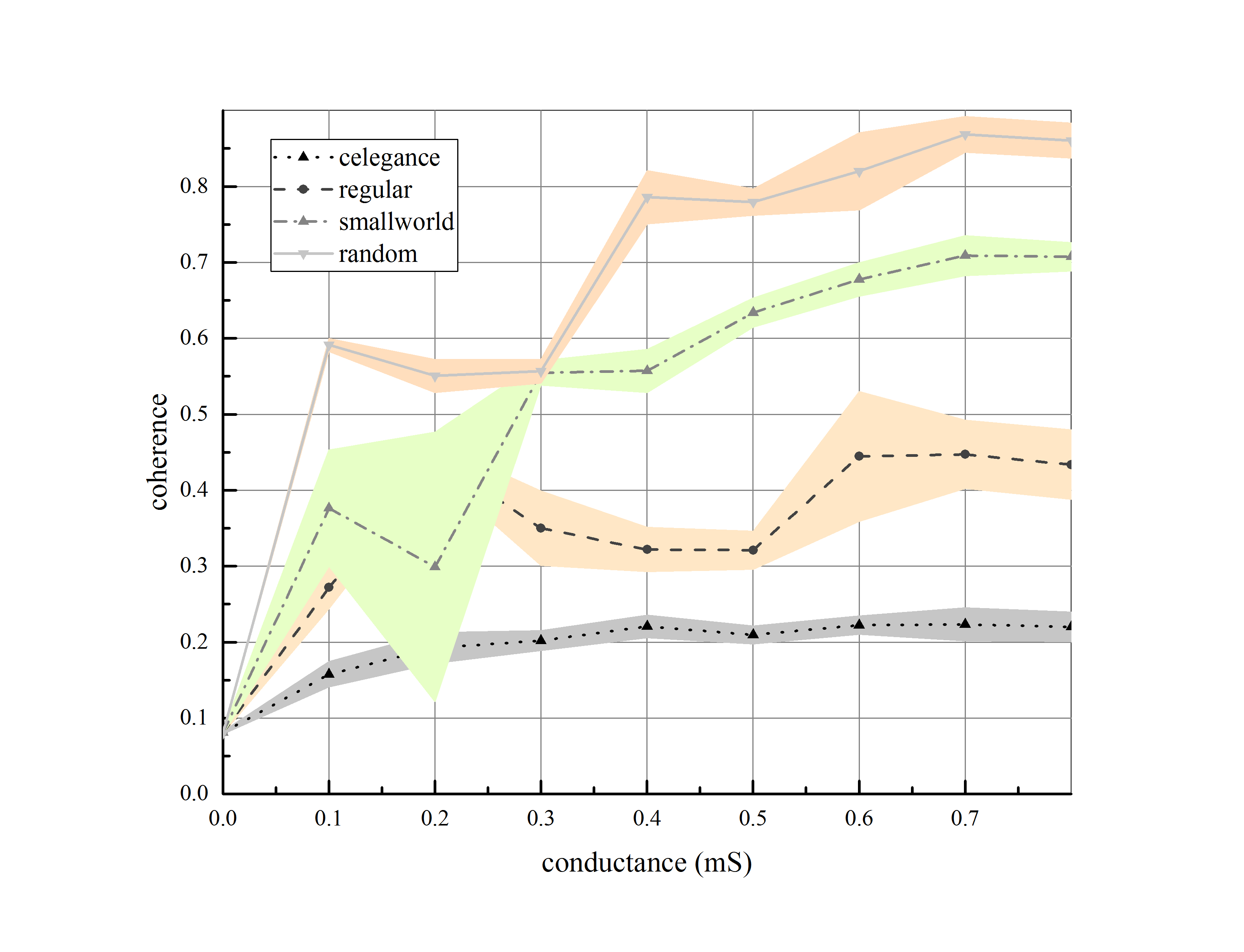}
\caption{A comparison of synchronization among C-elegance network, a regular network, a small-world network, and a random network for various amount of synaptic conductance. The synchronization index of C-elegance network has fewer amounts in comparison to other networks with almost the same average degree. Color margins show the standard error of experiment for five-time repetitions.}
\label{fig:8}
\end{figure}

C-elegance is one of the most studied micro-organisms, about 1mm in length, having a simple structure and a small number of nerve cells. Complete wiring map of C-elegance nervous system has been presented in various surveys \cite{WormAtlas}. In this study, we have utilized data reported in \cite{altunwormatlas} and investigated synchronization on C-elegance network. Considering chemical synapses, we have plotted the wiring diagram of C-elegance nervous system by Gephi software and then applied our model on it. There is a comparison of synchronization among C-elegance network and a regular network, a random network and a small-world network, with the same number of neurons and average connections for various amount of synaptic conductance in figure (\ref{fig:8}).

As illustrated, the coherence level of C-elegance network has fewer amounts in comparison to the other three networks with almost the same average degree. The topological structure of networks is compared and result has been reported in Table (\ref{tab4}).

\begin{table}[ht]
\centering
\begin{tabular}{|l|l|l|l|l|}
\hline
Structural Measure  & C-Elegance Network & Regular Network & Small-World Network & Random Network \\
\hline
Average Degree  & 15.728 & 48 & 48 & 48\\
\hline
Avg. Weighted Degree  & 22.918 & 24 & 24 & 24 \\
\hline
Network Diameter  & 10 & 12 & 4 & 3 \\
\hline
Graph Density  & 0.028 & 0.086 & 0.086 & 0.086 \\
\hline
Modularity  & 0.507 & 0.597 & 0.622 & 0.11 \\
\hline
Connected Components  & 1 & 1 & 1 & 1 \\
\hline
Avg. Clustering Coefficient  & 0.204 & 0.717 & 0.575 & 0.083 \\
\hline
Avg. Path Length  & 3.454 & 6.302 & 2.607 & 2.007 \\
\hline
\end{tabular}
\caption{\label{tab4} Calculating structural measures of C-elegance network, a regular network, a small-world network, and a random network. }
\end{table}

According to the structural measures in the Table (\ref{tab4}), the clustering coefficient of C-elegance network is bigger than the random network and the path length is shorter compared to the regular network. Therefore, we do not expect that kind of synchronization dynamic from C-elegance network. Consequently, we have selected a regular network with the same average degree as C-elegance’s and tried to close the structural parameters of both networks. Although differences in the structural measures are inconsiderable, there is a conspicuous difference in synchronization levels (see Table \ref{tab:5}).

\begin{table}[ht]
\centering
\begin{tabular}{|l|l|l|}
\hline
  & C-Elegance Network & WS- Network \\
\hline
Average Degree  & 15.728 & 15.943 \\
\hline
Avg. Weighted Degree  & 22.918 & 8 \\
\hline
Network Diameter  & 10 & 10  \\
\hline
Graph Density  & 0.028 & 0.029 \\
\hline
Modularity  & 0.507 & 0.552  \\
\hline
Connected Components  & 1 & 1 \\
\hline
Avg. Clustering Coefficient  & 0.204 & 0.279 \\
\hline
Avg. Path Length  & 3.454 & 4.227 \\
\hline
synchronization & 0.21105 & 0.46189 \\
\hline
\end{tabular}
\caption{\label{tab:5} A comparison of synchronization between C-elegance network and a regular network with the same structural measures. Although structural measures are similar, there is a conspicuous difference in synchronization levels. }
\end{table}

In order to find a logical reason for the different dynamics of C-elegance from other investigated topologies, we have defined two algorithms to decrease the clustering coefficient and the path length. By these reductions, the C-elegance structure would be similar to the random network. According to the decreasing algorithm of clustering coefficient, the clusters of each node (three-node motifs) are recognized, the third edge is eliminated and is randomly rewired to another vertex outside of that cluster. Moreover, in order to decrease the average path length, we have detected the shortest path from one vertex to all other vertices of the network and selected the longest one. Then, one of the edges is randomly eliminated and a path of unit length is wired between the first and the last vertices of the path. The result of applying these two algorithms has been reported in Table (\ref{tab:6}).

\begin{table}[ht]
\centering
\begin{tabular}{|l|l|l|l|}
\hline
  & C-Elegance Network & Decreasing Clustering Coefficient & Decreasing Path Length \\
\hline
Average Degree  & 15.728 & 15.355 & 15.735 \\
\hline
Avg. Weighted Degree  & 22.918 & 22.918 & 22.86\\
\hline
Network Diameter  & 10 & 6 & 9 \\
\hline
Graph Density  & 0.028 & 0.028 &0.028 \\
\hline
Modularity  & 0.507 & 0.388 & 0.511 \\
\hline
Connected Components  & 1 & 1 & 1 \\
\hline
Avg. Clustering Coefficient  & 0.204 & 0.02 & 0.199 \\
\hline
Avg. Path Length  & 3.454 & 3.036 & 3.411 \\
\hline
synchronization & 0.21105 & 0.10428 & 0.20216 \\
\hline
\end{tabular}
\caption{\label{tab:6}  Structural measures and synchronization indexes of C-elegance after and before applying decreasing algorithm of clustering coefficient and path length.}
\end{table}

\begin{figure}[H]
  \centering
\includegraphics[scale=0.55]{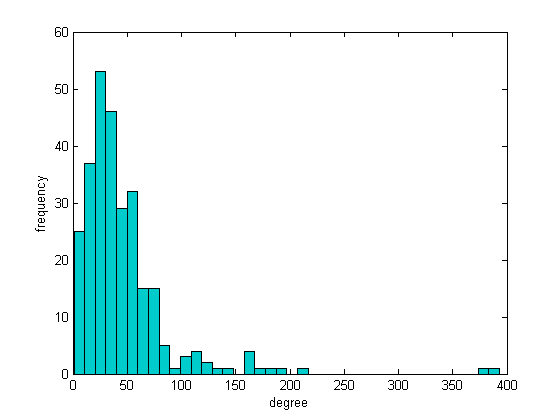}
\includegraphics[scale=0.55]{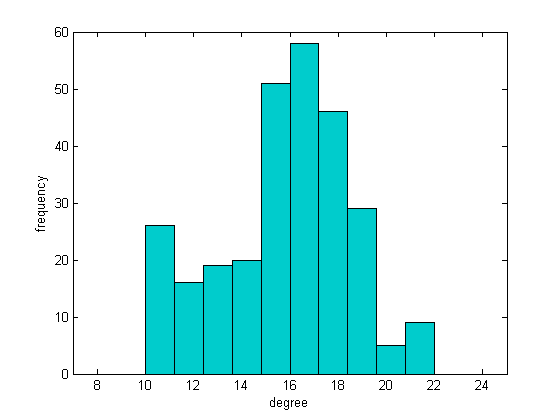}
\caption{The degree distribution of the C-elegance network (left) and a WS network with a similar structural parameters to C-elegance (right).}
\label{fig:9}
\end{figure}

We expected an increasing trend of synchronization by applying these algorithms because the structural measures are getting near to random network, but the result was exactly vice versa. According to Table (\ref{tab:6}), there is a reduction in synchronization of this exchanged networks. In this way, we have plotted degree distributions of two networks having similar structural parameters (Table (\ref{tab:5})). As shown in figure (\ref{fig:9}), there is a noticeable difference between results, so it makes sense we deduce that another important parameter playing an effective role in synchronization index, and it could be the degree distribution.

\section{Conclusion}
Investigation of the relation among topology and dynamical processes are of great importance in neuroscience because it provides scholars with a better understanding of the brain through leading them towards a solution to the issue of the connection between the structure and the function of networks. In this paper, we have investigated the synchronization dynamic on neuronal networks. First, through simulation, we have measured the coherence level of neuronal networks with different topologies, from a regular network to a small-world and then to a random network. It is worth noting that the Hodgkin-Huxley model, defined as neurons’ dynamic and excitatory and inhibitory synaptic currents, considered with AMPA receptors and $GABA_A$, respectively. Considering a completely excitatory network without any delay in transferring synaptic currents, the least amount of synchronization has been associated with a regular network with synchronous clusters. There is a sharp increase in the coherence level as the network topology changes to a small-world. This increasing trend grows steadily, and the maximum level of synchronization is experienced by a random network.

To determine the role of excitatory currents in synchronization, we have exchanged the percentages of excitatory neurons for inhibitory neurons. These changes have a negative effect on synchronization, and there is an almost zero level of synchronization in a completely inhibitory network. Furthermore, the topology is not effective in this inhibitory network, so through some assumptions, it is shown that the excitatory currents play a crucial role in synchronization.

As an important step in simulation, we have tried to find and examine crucial factors of the network structure influenced by synchronization. The number of neurons, the average number of connections and the amount of synaptic conductance are the parameters investigated in this paper. The last part of our research has been allocated to the utilization of the model to a C-elegance network, as an available network. Despite the fact that the structural measures of this network are in the range of networks having been defined so far, the dynamic of the network is completely different. Thus, we have deduced that there should be another factor influenced by synchronization, and we guess it is the degree distribution.

\bibliography{sample}

\end{document}